\documentstyle[11pt]{article}

\begin{document}
\begin{center}
{\Large What was the fastest 100m final?} \\
\vspace{5mm}
J. R. Mureika \\
{\it Department of Computer Science \\
University of Southern California \\
Los Angeles, CA~~90089-2520}
\end{center}

Over the past 10 years, we've seen a number of exceptional sprint
performances at the world class level.  Of particular note has been the
recent emergence of a handful of contenders to the title of ``World's Fastest 
Man''.  A decade previous, we would have been talking Ben Johnson and Carl 
Lewis.  Today, we find it hard to choose between Donovan Bailey, Frank 
Fredericks, Ato Boldon, and Maurice Greene.  Of course, we can't forget
(recently stepped-down) European Sprint King Linford Christie, or former
WR holder Leroy Burrell.  The  increase in number of top world class
sprinters makes one wonder: what really was the fastest 100m final?\\

Recently, I've written several articles which have discussed correcting for
wind effects in the sprints \cite{me1,me2}: this provides an easy way to compare most 100m
times run in essentially any wind condition (head-  or tail-wind).
Unfortunately the model doesn't take temperature into account, but
susceptibility to temperature must almost certainly be an individual factor
(proof: I find 15$^o$ weather pleasant and cool, while all other Californians
don their parkas).  The findings have been most interesting, and have
re-written the record books to a certain degree.\\

In the following article, I will analyze the results from key races over the
past 10 years.  These include the 1983-1997 World Championships (WC) finals,
1984-1996 Olympic Games (OG) finals, and the 1996 Lausanne Grand Prix (LGP)
final, in which Frank Fredericks ran his 9.86s PB into a 0.4 m/s headwind.
The latter race was considered by many to be the finest 100m performance
ever.  In fact, after wind-correction, this adjusts
to a calm 9.84s, numerically matching Donovan Bailey's 9.84 (+0.7 m/s).
Meanwhile, Bailey's WR mark translates to a 9.88s, and is usurped by 
Fredericks' Lausanne run (Table~\ref{table1}). \\

\noindent{\bf Fastest average race times} \\

First, let's look at the average wind-corrected times for each final
(Table~\ref{fastestfinal}).  The only race with a sub-10s
average time is the 1996 LGP.  This low value is
weighted by the fact that 4 of the competitors clocked sub-10s runs,
and 4th-7th place were within 0.01s of each other.  The WR race in Atlanta
is ranked 3rd in terms of average time, and surprisingly the 1991 WC
final (which featured the most legal sub-10s runs in a single race) ranks
only 4th.  In fact, the average times of the 1996 OG and 1991 WC finals
are almost identical (after wind-correction).  Near the bottom of the list
are the 1983 WC and 1984 OG, which featured few spectacular
performances (retroactively speaking).  That is, with the exception of Carl
Lewis, the times were quite far off the WR of the time (Calvin Smith's 
9.93s from Colorado Springs, 07 Jul 1983). \\

Note, however, that even though they were monumental races of their day,
the 1987 WC and 1988 OG runs are ranked last!  This is due to the fact that
the last place finishers in each race (Pierfrancesco Pavoni ITA - 16.38,
Ray Stewart JAM - 12.26) were obviously not running at peak potential.
This raises a serious concern when dealing with the {\it overall}
average times of races.   \\

It seems logical that one can divide a race into
two distinct groups: the top 4 or 5 finishers are most likely
the serious medal contenders, and can be taken to represent the ``quality''
of the final.  Those who finish 5th-8th do so for any number 
of reasons, which might include: (a) they don't match the calibre of the top 
finalists, and finish at the best of their ability, (b) they pull up in the 
race or are running injured, or (c) they shut down before the finish line 
because of mental duress ({\it e.g.} Merlene Ottey at the 1997 WC). \\

\noindent{\bf The average winning margin}\\

Since the overall average time for a final can apparently be misleading, as
per the conclusion of the previous section, it might make more sense to
consider the {\it average winning margins} of each race.  That is,
by how much, on average, did the gold medallist defeat his competitors?
Adhering to the KISS principle (Keep It Statistically Simple, not developed
by Gene Simmons), there is a straightforward expression for calculating this
beast:

\begin{equation}
^{\rm Average~winning}_{\rm margin}~ = ~\frac{k}{k-1}~ \left(
^{\rm Average~time~of}_{{\rm first~k~competitors}}~~ - ~~
^{\rm Winning}_{\rm time} \right)~.
\end{equation}

The winning margin can help up compare top finishing places in different races,
and can provide more information besides just how far ahead of the rest
of the field was the gold medalist. \\

While the results from each race are wind-corrected to provide an easy 
ground for comparison, the quantity above is essentially independent of this
correction.  In a typical world class race (where the first and last place
times fall within about a 0.3s interval), wind-correction roughly amounts
to a shift of an overall constant (a couple hundredths of a second), and
this overall constant cancels out in the above equation (trust me!).  So,
the average winning margin as calculated by this method will be the same
regardless of whether or not the times are wind-corrected or official.\\

Tables~\ref{top3},\ref{top4} rank the average finishing times for the top 3
and 4 competitors, while Tables~\ref{wm3},\ref{wm4} order the winning 
margins in increasing order.  This gives a sense of the ``closeness'' of the 
race: the smaller the average winning margin, the closer the finish, and the
closer the calibre of the athletes in the final.  \\

While the increase from 3 to 4 finishers rearranges the lists, 
there are several 
key constants of note.  In each case (Tables~\ref{top3},\ref{top4}),
the fastest average race is the 1996 Lausanne Grand Prix, the wind-corrected
World Record race which produced Fredericks' headwind 9.86 dash.
The Atlanta final (1996 OG) takes 2nd and 3rd place in the averages 
rankings, trailing the 96 LGP average finishes by roughly 0.02s.    
It posts the 2nd smallest winning margin for top 3, but slips to to
4th for the top 4 finalists. \\

The 1991 WC final holds its ranking as the ``closest'' race, yielding
a winning margin of 0.035s and 0.044s for top 3 and 4 placings.  As noted
earlier, this race posted the most sub-10s marks ever, but after a
wind-correction treatment, the average top 3 and 4 times rank only
5th and 4th, respectively.  This is a good example of how a race can
seem faster than it really is because of tailwind effects, but even
after correction can still be considered quite an impressive sprint!\\

Conversely, the races with the lowest average for top 3 and 4 are the
1984 OG and 1983 and 1987 WCs.  Likewise, these represent the largest average
winning margins.  This information is quite useful: in each race, Carl 
Lewis was the clear winner.  Apparently he was quite ahead of his time!
Interestingly enough, Lewis is also a factor in the 1991 WC race, posting
his legal (and new WR) 9.86s jolt.\\

\noindent{\bf The 1987 WC and 1988 OG} \\

As we all know too well, the 1987 WC and 1988 OG finals were particularly out
of the ordinary.  Ben Johnson's then-WR marks of 9.83s and 9.79s were
themselves about 10 years ahead of their time, having been only recently 
clocked by other world class contenders after wind-correction.  The 9.83s
mark (+1.0 m/s) corresponds to a 9.89s still-air run, which has been matched
or bettered by several athletes in recent years (Bailey, Greene at 9.88s,
and Bailey, Christie, Burrell at 9.89s).  Meanwhile, the 9.79s (+1.1) adjusts 
to 9.85s, having only been topped by Fredericks' infamous 1996 LGP mark. \\

If we were to consider the results to be official, how would the findings
be affected?  Table~\ref{ben} shows the appropriate statistics for the
races in question, including Johnson's stricken marks.  Again, the overall
averages are unusually high, due to the lackluster clockings of 8th place
(10.948s and 10.346s).  Without last place, the overall averages lower to
10.193s and 10.059s, ranking these after the 1997 WC and 1995 WC.   \\

Despite the anomalous last place times, the winning margin considered earlier
is unaffected by these.  For the Johnson races, we have the average top 3
times of 10.007s and 9.970s, with similar respective winning margins of 0.176s 
and 0.180s.  According to this, the Seoul final would rank 6th for top 3
finishers, and would mark one of the largest winning margins for the races
considered.  In terms of top 4 finishers, we have 10.055s and 9.985s, with
winning margins of  0.220s and 0.180s.  Again, these constitute some of
the largest winning margins, and help to show that Ben Johnson's performances
were well in advance of the rest of the world (if we ignore {\it why} they
were at such a level).  \\

\noindent{\bf Winning margins and World Record progression} \\

Another interesting way to judge the ``calibre'' of a 100m final is obviously
to compare it to the current WR performance.  The long/triple jump always
have a WR mark at the side of the pits, so the spectators can get an idea of
how close/far the competitor was from the crown.  Why not have one in the
sprints?   \\

This can roughly be done in a similar manner to the way that the winning margin
was calculated earlier.  The question asked is: how far behind the WR were the
top k competitors in this race?  Along with the winning margins, 
Tables~\ref{top3} and \ref{top4} list the deviation of the top 3 and 4 
average times from the WR (see also Table~\ref{wr}).   \\

For the cases where the WR was set in the race, the winning margins and the 
deviation from the WR are the same, since they're calculated in exactly the same
manner.  For the other cases, the method is slightly modified.  Instead of
using the earlier expression involving the fraction $k/(k-1)$, we just 
simply subtract the WR time from the average of the top 3 and 4.   \\

Note a slight difference here.  In order to obtain this quantity with a 
minimal amount of work, I have {\it not} wind-corrected these quantities.
Had I done this, I would have spent quite a while going back through the
record books to find the wind-corrected WRs of each year, since we know
that a great performance can be masked by a suitably strong head-wind
({\it e.g.} Bailey's 10.03s in Abbotsford earlier this year \cite{me2}).  So, to 
make the numbers more ``useful'' to the naked eye, the wind conditions are
not performed, which can skew the data a bit when the wind conditions
for the WR race are sufficiently different for the race in question 
(case and point: the 1996 LGP winning margin {\it v.s.} the deviation from
the 9.85 WR). \\

\noindent{\bf Musings} \\

All this being said and done, how can we answer the question at hand?  Which
race really is the fastest 100m?  The easiest answer is that ther is no
definite answer.  It all depends on what is meant by the {\it fastest}.
Here are some points to consider:

\begin{itemize}
\item{The 1996 LGP produced the fastest wind-corrected time ever (9.84s),
and as a partial result yields the smallest average times (including overall,
top 3, and top 4 finishers)}
\item{The 1991 WC race, while not seeming so fast after wind-correction,
was the closest race of all those considered, and could be understood to have
had the most on-equal-par athletes competing}
\item{Carl Lewis' performances in the 1980s were several years ahead of
their time, putting him far above the competition of the time}
\end{itemize}

The results would tend to suggest that there are more exceptional
World Class athletes today who are of equal calibre than ever before.  
It's interesting to think what these lists might look like in another
ten years!\\

\vskip .25 cm
 
\noindent
{\bf Acknowledgements}

I thank Roman Mureika (Department of Mathematics and Statistics, University
of New Brunswick, Canada) for helpful suggestions about the statistical
analysis.  I also thank Cecil Smith (Ontario Track and Field Association, 
Canada) for providing
the wind reading for the 1984 Olympic Games in Los Angeles.  Most results
herein are from the Track and Field Statistics Website,
{\tt http://www.uta.fi/$\sim$csmipe/sport/index.html}, or from the
International Amateur Athletics Federation (IAAF) site, 
{\tt http://www.iaaf.org/}.

\pagebreak

\begin{table}[t]
\begin{center}
{\begin{tabular}{|l l l l|}\hline
1. Frank Fredericks & 9.84 (9.86, -0.4) & Lausanne GP &  1996 \\
2. Donovan Bailey   & 9.88 (9.84, +0.7) & Atlanta OG  &  1996 \\
3. Maurice Greene   & 9.88 (9.86, +0.2) & Athens WC   &  1997 \\ \hline
\end{tabular}}
\end{center}
\caption{Top 3 fastest indivuduals (as of 01 Sep 1997)}	
\label{table1}
\end{table}

\begin{table}[h]
\begin{center}
{\begin{tabular}{|l l|}\hline
1. 1996 LGP &9.981s \\
2. 1993 WC &10.029 \\
3. 1996 OG &10.033 \\
4. 1991 WC &10.038 \\
5. 1997 WC &10.044 \\
6. 1992 OG &10.133 \\
7. 1995 WC &10.161 \\
8. 1983 WC &10.249 \\
9. 1984 OG &10.256 \\
10. 1988 OG &10.417 \\
11. 1987 WC &11.100 \\ \hline
\end{tabular}}
\end{center}
\caption{Fastest Races (average, all competitors)}
\label{fastestfinal}
\end{table}
 
\begin{table}[h]
\begin{center}
{\begin{tabular}{|l l l|}\hline
1. 1996 LGP&9.890&(1. Fredericks 9.84, 2. Bailey 9.91, 3. Boldon 9.92)  \\
2. 1996 OG&9.917&(1. Bailey 9.88, 2. Fredericks 9.93, 3. Boldon 9.94) \\
3. 1997 WC&9.923&(1. Greene 9.88, 2. Bailey 9.93, 3. Montgomery 9.96) \\
4. 1993 WC&9.947&(1. Christie 9.89, 2. Cason 9.94, 3. Mitchell 10.01) \\
5. 1991 WC&9.953&(1. Lewis 9.93, 2. Burrell 9.95, 3. Mitchell 9.98) \\
6. 1988 OG&10.030& (1. Lewis 9.99, 2. Christie 10.04, 3. Smith 10.06) \\
7. 1992 OG&10.037&(1. Christie 9.99, 2. Fredericks 10.05, 3. Mitchell 10.07) \\
8. 1995 WC&10.070&(1. Bailey 10.03, 2. Surin 10.09, 3. Boldon 10.09) \\
9. 1987 WC&10.110&(1. Lewis 9.99, 2. Stewart 10.14, 3. Christie 10.20) \\
10. 1984 OG&10.153&(1. Lewis 10.01, 2. Graddy 10.21, 3. Johnson 10.24) \\
11. 1983 WC&10.249&(1. Lewis 10.06, 2. Smith 10.20, 3. King 10.23) \\ \hline
\end{tabular}}
\end{center}
\caption{Fastest Top 3 Finishers (average)}
\label{top3}
\end{table}

\begin{table}[h]
\begin{center}
{\begin{tabular}{|l c c|}\hline
 & Winning  & Deviation \\
 & margin  & from WR \\ \hline
1. 1991 WC&0.035 & 0.035 \\
2. 1996 OG&0.056 & 0.056 \\
3. 1995 WC&0.060 & 0.160\\
4. 1997 WC&0.065 & 0.063 \\
5. 1992 OG&0.071 & 0.147 \\
6. 1996 LGP&0.075 & 0.060 \\
7. 1988 OG&0.080 & 0.060 \\
8. 1993 WC&0.086 & 0.067 \\
9. 1983 WC&0.155 & 0.243 \\
10. 1987 WC&0.180 & 0.180 \\
11. 1984 OG&0.215 & 0.203 \\ \hline
\end{tabular}}
\end{center}
\caption{Smallest average winning margins and deviation from WR, top 3 }
\label{wm3}
\end{table}
 
\begin{table}[h]
\begin{center}
{\begin{tabular}{|l l l|}\hline
1. 1996 LGP&9.913&(4. Drummond 9.98) \\
2. 1997 WC&9.935&(4. Fredericks 9.97) \\
3. 1996 OG&9.945&(4. Mitchell 10.03) \\
4. 1991 WC&9.963&(4. Christie 9.99) \\
5. 1993 WC&9.970&(4. Lewis 10.04) \\
6. 1988 OG&10.050&(4. Mitchell 10.11) \\
7. 1992 OG&10.058&(4. Surin 10.12) \\
8. 1995 WC&10.085&(4. Fredericks 10.13) \\
9. 1987 WC&10.148&(4. Kovacs 10.26) \\
10. 1984 OG&10.185&(4. Brown 10.28) \\
11. 1983 WC&10.188 &(4. Wells 10.26) \\ \hline
\end{tabular}}
\end{center}
\caption{Fastest Top 4 Finishers (average)}
\label{top4}
\end{table}

\begin{table}[h]
\begin{center}
{\begin{tabular}{|l c c|}\hline
 & Winning  & Deviation \\
 & margin  & from WR \\ \hline
1. 1991 WC&0.044 & 0.044 \\
2. 1997 WC&0.073 & 0.075 \\
   1995 WC&0.073 & 0.175 \\
4. 1988 OG&0.080 & 0.080 \\
5. 1996 OG&0.087 & 0.087 \\
6. 1992 OG&0.091 & 0.168 \\
7. 1996 LGP&0.097 & 0.083 \\
8. 1993 WC&0.107 & 0.090 \\
9. 1983 WC&0.171 & 0.268 \\
10. 1987 WC&0.211 & 0.211 \\
11. 1984 OG&0.233 & 0.235 \\ \hline
\end{tabular}}
\end{center}
\caption{Smallest average winning margins and deviation from WR, top 4}
\label{wm4}
\end{table}

\begin{table}[h]
\begin{center}
{\begin{tabular}{|l | l l| l l| l l|}\hline
Race & Overall& Average & Average & winning  & Average  & winning \\
 & average & (no 8th) & top 3 & margin & top 4 & margin \\  \hline
1987 WC & 10.948 & 10.193 & 10.007 & 0.176 & 10.055 &  0.220 \\
1988 OG & 10.417 & 10.059 & 9.970 & 0.180 & 9.985 & 0.180 \\ \hline
\end{tabular}}
\end{center}
\caption{Stats for 1987 WC and 1988 OG, including Ben Johnson's performances}
\label{ben}
\end{table}

\begin{table}[h]
\begin{center}
{\begin{tabular}{|l l l l|}\hline
WR & Athlete & Date & Location \\ \hline
9.93A (+1.4) &    Calvin Smith           USA &03 Jul 1983 &Colorado Springs \\ 
9.93 (+1.0) &    Carl Lewis             USA &30 Aug 1987 &Rome \\ 
9.92 (+1.1) &    Carl Lewis             USA &24 Sep 1988 &Seoul \\ 
9.90 (+1.9) &    Leroy Burrell          USA &14 Jun 1991 &New York \\ 
9.86 (+1.2) &    Carl Lewis             USA &25 Aug 1991 &Tokyo \\ 
9.85 (+1.2) &    Leroy Burrell          USA &06 Jul 1994 &Lausanne \\ 
9.84 (+0.7) &    Donovan Bailey         CAN &27 Jul 1996 &Atlanta \\  \hline
\end{tabular}}
\end{center}
\caption{100m world record progression, 1983 - 1997.}
\label{wr}
\end{table}

\begin{table}[h]
\begin{center}
{\begin{tabular}{|l l l|}\hline
Athlete	& Official & Wind-corrected \\ \hline
\multicolumn{3}{|l|} {\bf 1997 WC, Athens (wind +0.2 m/s)} \\ 
1. Maurice Greene (USA)   &     9.86  &  9.88  \\
2. Donovan Bailey (CAN)   &     9.91  &  9.93  \\
3. Tim Montgomery (USA)   &     9.94  &  9.96  \\
4. Frank Fredericks (NAM) &     9.95  &  9.97  \\
5. Ato Boldon (TRI)       &     10.02 &  10.04 \\
6. Davidson Ezinwa (NIG)  &     10.10 &  10.12 \\
7. Bruny Surin (CAN)      &     10.12 &  10.14 \\
8. Mike Marsh (USA)       &     10.29 &  10.31 \\ \hline \hline
\multicolumn{3}{|l|} {\bf 1995 WC, Goetenburg (+1.0)} \\
1. Donovan Bailey (CAN)	&9.97&	10.03   \\
2. Bruny Surin (CAN)    &10.03&	10.09   \\
3. Ato Boldon (TRI)     &10.03&	10.09   \\
4. Frank Fredericks (NAM)& 10.07&	10.13 \\
5. Michael Marsh (USA)    &10.10 &	10.16 \\
6. Linford Christie (GBR) & 10.12&	10.18 \\
7. Olapade Adeniken (NGR)& 10.20&	10.26 \\
8. Raymond Stewart (JAM) & 10.29&	10.35 \\ \hline \hline
\multicolumn{3}{|l|} {\bf 1993 WC, Stuttgart (+0.3)} \\
1. Linford Christie (GBR) &	9.87  &	9.89   \\
2. Andre Cason (USA)     &9.92  &	9.94   \\
3. Dennis Mitchell (USA)  &  	9.99  &	10.01  \\
4. Carl Lewis (USA)     	&10.02 & 10.04  \\
5. Bruny Surin (CAN)     &10.02 &	10.04  \\
6. Frank Fredericks (NAM)&    	10.03 &	10.05  \\
7. Daniel Effiong (NGR)  &  	10.04 &	10.06  \\
8. Raymond Stewart (JAM) &   	10.18 &	10.20  \\\hline \hline
\multicolumn{3}{|l|} {\bf 1991 WC,  Tokyo (+1.2)} \\
1. Carl Lewis (USA)     &	9.86 &	9.93   \\
2. Leroy Burrell (USA)  &  	9.88 &	9.95   \\
3. Dennis Mitchell (USA)&    	9.91 &	9.98   \\
4. Linford Christie (GBR) &   	9.92 &	9.99   \\
5. Frank Fredericks (NAM)&    	9.95 &	10.02  \\
6. Ray Stewart (JAM)     & 9.96 & 	10.03 \\
7. Robson da Silva (BRA) &   	10.12& 	10.19 \\
8. Bruny Surin (CAN)     &10.14& 	10.21 \\ \hline \hline
\end{tabular}}
\end{center}
\end{table}
\pagebreak
\begin{table}[h]
\begin{center}
{\begin{tabular}{|l l l|}\hline
\multicolumn{3}{|l|} {\bf 1987 WC, Rome (+1.0)} \\
DQ. Ben Johnson (CAN) &  9.83 & 9.89   \\
1. Carl Lewis (USA)     &9.93 & 	9.99 \\
2. Raymond Stewart (JAM) &  10.08	&10.14  \\
3. Linford Christie (GBR) &  10.14	&10.20  \\
4. Attila Kovacs (HUN)&  10.20	&10.26  \\
5. Viktor Bryzgin (USR)   &  10.25	&10.31  \\
6. Lee McRae (USA)     	&10.34&	10.41 \\	
7. Pierfrancesco Pavoni (ITA) &  16.23 &16.38 \\ \hline \hline
\multicolumn{3}{|l|} {\bf  1983 WC, Helsinki (-0.3)} \\
1. Carl Lewis (US)     	&10.07	&10.06   \\
2. Calvin Smith (US)     &10.21	&10.20	 \\
3. Emmit King (US)     	&10.24	&10.23	 \\
4. Allan Wells (GB)     &10.27	&10.26	 \\
5. Juan Núñez (DR)    &	10.29&	10.28	 \\
6. Christian Haas (WG)     &10.32&	10.31  \\
7. Paul Narracott (Aus)     &10.33&	10.31  \\
8. Desai Williams (Can)     &10.36&	10.34  \\ \hline \hline
\multicolumn{3}{|l|} {\bf 1996  LGP, Lausanne (-0.4)} \\
1. Frank Fredericks (NAM) & 	9.86 &	9.84 \\
2. Donavan Bailey (CAN)   & 	9.93&	9.91 \\
3. Ato Boldon (TRI)       & 	9.94&	9.92 \\
4. Jon Drummond (USA)     &	10.00&	9.98 \\
5. Linford Christie (GBR) &	10.04&	10.02\\
6. Bruny Surin (CAN)      &	10.05&	10.03\\
7. Leroy Burrel (USA)     &	10.05&	10.03\\
8. Dennis Mitchell (USA)  &	10.15&	10.12\\ \hline \hline
\multicolumn{3}{|l|} {\bf 1996 OG, Atlanta (+0.7)} \\
1. Donovan Bailey   (CAN) &   9.84  & 9.88   \\ 
2. Frank Fredericks (NAM) &   9.89  & 9.93   \\
3. Ato Boldon       (TRI) &   9.90  & 9.94   \\
4. Dennis Mitchell  (USA) &   9.99  & 10.03  \\
5. Mike Marsh       (USA) &   10.00 & 10.05  \\
6. Davidson Ezinwa  (NGR) &   10.14 & 10.19  \\
7. Michael Green    (JAM) &   10.16 & 10.21  \\ 
8. Linford Christie (GBR) &   DQ    &     \\ \hline \hline
\end{tabular}}
\end{center}
\end{table}
\pagebreak
\begin{table}[h]
\begin{center}
{\begin{tabular}{|l l l|}\hline
\multicolumn{3}{|l|} {\bf 1992 OG, Barcelona (+0.5)} \\
1. Linford Christie       (GBR)&  9.96 	&9.99 \\
2. Frankie Fredericks     (NAM)& 10.02 	&10.05 \\
3. Dennis Mitchell        (USA)& 10.04 	&10.07 \\
4. Bruny Surin            (CAN)& 10.09 	&10.12 \\
5. Leroy Burrell          (USA)& 10.10 	&10.13 \\
6. Olapade Adeniken       (NGR)& 10.12 	&10.15 \\
7. Ray Stewart            (JAM)& 10.22 	&10.25 \\
8. Davidson Ezinwa        (NGR)& 10.26 	&10.30  \\ \hline
\multicolumn{3}{|l|} {\bf 1988 OG, Seoul (+1.1)} \\
DQ.   Ben Johnson     (CAN)  &9.79   &	 9.85 \\
1. Carl Lewis             (USA)& 9.92 &	9.99  \\
2. Linford Christie       (GBR)& 9.97 &	10.04 \\
3. Calvin Smith           (USA)& 9.99 &	10.06 \\
4. Dennis Mitchell        (USA)&10.04 &	10.11 \\
5. Robson da Silva        (BRA)&10.11 &	10.18 \\
6. Desai Williams         (CAN)&10.11 &	10.18 \\
7. Ray Stewart            (JAM)&12.26 &	12.36 \\ \hline \hline
\multicolumn{3}{|l|} {\bf 1984 OG, Los Angeles (+0.2)} \\
1. Carl Lewis             (USA) & 9.99&	10.01 \\
2. Sam Graddy             (USA) &10.19&	10.21 \\
3. Ben Johnson            (CAN) &10.22&	10.24 \\
4. Ron Brown              (USA) &10.26&	10.28 \\
5. Mike McFarlene         (GBR) &10.27&	10.29 \\
6. Ray Stewart            (JAM) &10.29&	10.31 \\
7. Donovan Reid           (GBR) &10.33&	10.34 \\
8. Tony Sharpe            (CAN) &10.35&	10.37 \\ \hline
\end{tabular}}
\end{center}
\caption{Major competition results, 1983-1997}
\end{table}

\end{document}